\documentclass[aps,pre,twocolumn,showpacs,preprintnumbers,amsmath,amssymb]{revtex4}

\usepackage{graphicx}
\usepackage{dcolumn}
\usepackage{bm}

\begin{document}

\title{Nonlinear evolution of the plasma beatwave: Compressing the laser beatnotes
via electromagnetic cascading}
\author{Serguei Kalmykov}\email{kalmykov@physics.utexas.edu} \author{Gennady Shvets}
 \affiliation{Department of Physics and Institute for
Fusion Studies, The University of Texas at Austin,  One University
Station C1500, Austin, Texas 78712}
\date{\today}

\begin{abstract}
The near-resonant beatwave excitation of an electron plasma wave
(EPW) can be employed for generating the trains of few-femtosecond
electromagnetic (EM) pulses in rarefied plasmas. The EPW produces
a co-moving index grating that induces a laser phase modulation at
the difference frequency. The bandwidth of the phase-modulated
laser is proportional to the product of the plasma length, laser
wavelength, and amplitude of the electron density perturbation.
The laser spectrum is composed of a cascade of red and blue
sidebands shifted by integer multiples of the beat frequency. When
the beat frequency is lower than the electron plasma frequency,
the red-shifted spectral components are advanced in time with
respect to the blue-shifted ones near the center of each laser
beatnote. The group velocity dispersion of plasma compresses so
chirped beatnotes to a few-laser-cycle duration thus creating a
train of sharp EM spikes with the beat periodicity. Depending on
the plasma and laser parameters, chirping and compression can be
implemented either concurrently in the same, or sequentially in
different plasmas. Evolution of the laser beatwave end electron
density perturbations is described in time and one spatial
dimension in a weakly relativistic approximation. Using the
compression effect, we demonstrate that the relativistic
bi-stability regime of the EPW excitation [G. Shvets, Phys. Rev.
Lett. {\bf 93}, 195004 (2004)] can be achieved with the initially
sub-threshold beatwave pulse.
\end{abstract}


\maketitle


\section{Introduction}
An electron plasma wave (EPW) is a natural tool for manipulating
the properties of intense radiation beams. It can be used for
up-shifting the laser frequency~\cite{Upshifting}, for enhancing
the self-focusing of co-propagating~\cite{Gibbon1,Gibbon2} and
counter-propagating~\cite{ShvetsPukhov} radiation beams, for the
resonant self-modulation of the laser amplitude~\cite{Andreev},
and for coupling the signal and pump lasers in the parametric
amplifier~\cite{Amplification}. Also, the high-amplitude EPW
driven by a short laser pulse can induce the pulse shrinkage with
time~\cite{Faure}.

Excitation of the EPW by the ponderomotive force (beatwave) of the
two-color laser with the difference frequency $\Omega$ close to
the electron Langmiur frequency $\omega_p=\sqrt{4\pi e^2 n_0/m_e}$
($n_0$ is an electron plasma density, $m_e$ and $-|e|$ are the
electron rest mass and charge) has attracted attention for a long
time~\cite{rosenbluth_liu,Tajima,Tang,Tochitski}. The nonlinear
plasma wave is highly sensitive to the variations of frequency and
amplitude of the ponderomotive force. For example, by chirping the
beat frequency~\cite{Lindberg} the EPW excitation can be enhanced
by the autoresonance effect. Downshifting the beat frequency from
the plasma resonance ($\Omega<\omega_p$) can also result in the
large-amplitude wake excitation due to the effect of relativistic
bi-stability~\cite{bistability,Royal}. Conversely, the driven
electron density perturbations can cause the laser amplitude
modulation, either transverse~\cite{Gibbon1,Gibbon2} or
longitudinal~\cite{KS}. Therefore, the performance of the beatwave
scheme critically depends on the self-consistent evolution of the
light and plasma waves that includes effects of numerous plasma
nonlinearities~\cite{Mori97}. The relativistic self-phase
modulation~\cite{Watts}, stimulated forward Raman
scattering~\cite{Mori,Sakharov} (SFRS), and electromagnetic
cascading~\cite{Gibbon1,Gibbon2,KS,Kaufman,Salomaa} (EMC) broaden
the laser frequency bandwidth, while the group velocity dispersion
(GVD) of radiation distorts the laser amplitude. In certain
regimes this scenario results in a strong local enhancement of the
laser field. Specifically, the frequency downshifted
($\Omega<\omega_p$) beatwave pulse of initially low amplitude can
be transformed into a train of sharp electromagnetic (EM) spikes
of relativistic intensity and few-laser-cycles duration~\cite{KS}.
The spikes are separated in time by the beat period
$\tau_b=2\pi/\Omega$.

In the present paper a theoretical model is formulated which, in a
weakly relativistic approximation, accurately describes the
nonlinear excitation and relativistic bi-stability of the EPW, and
frequency and amplitude modulation of the laser pulse in one
spatial dimension (1D) and in time. We select specific regimes in
which the EMC induced by near-resonantly driven EPW causes laser
spectral broadening. At every point of the perturbed plasma, the
EPW creates an index grating co-moving with the laser beams.
Hence, the periodic frequency modulation (FM) of the laser
develops at a difference frequency $\Omega$. In spectral terms,
the FM is manifested as a cascade of Stokes and anti-Stokes
sidebands shifted by integer multiples of $\Omega$ from the laser
fundamental $\omega_0$.

The effect of EMC has been known in physics of laser-plasma
interactions since early 70s when Cohen {\it et
al.}~\cite{Kaufman} suggested to enhance plasma heating by the
decay of the cascade-driven EPW. Later on, the EMC was considered
as an EPW diagnostic in the plasma beatwave
accelerator~\cite{Salomaa}. Systematic
study~\cite{Gibbon2,Gibbon1} of plasma waveguiding options
provided by the nonlinear interaction of laser beams with the
cascade-driven EPW revealed an enhanced self-focusing of the
co-propagating beams detuned in frequency below plasma resonance
($\Omega<\omega_p$). Calculations of Refs.~\cite{Gibbon2,Gibbon1}
describe the non-stationary cascade evolution in two dimensions
(2D) in the planar and three dimensions (3D) in cylindrical
geometry, take a full account of the relativistic nonlinearities
of both cascade components and the EPW, but neglect the GVD of
radiation and, hence, the longitudinal (temporal) distortion of
the laser amplitude caused by the longitudinal transport of the
cascade energy. Our work fills this gap by concentrating on the 1D
compression of laser beatnotes due to the EMC and GVD.

\begin{figure}[t]
\includegraphics[scale=1]{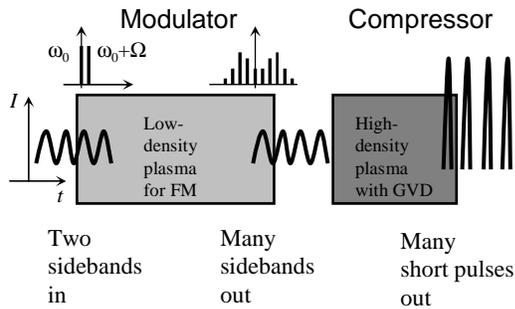}
\caption{\label{Fig1} Schematic of the two-stage cascade
compressor. The frequency modulation (FM) occurs in a rarefied
plasma. Denser plasma is used for the compression of beatnotes.}
\end{figure}

%

We prove that the GVD completely dominates the evolution of
weakly-relativistic beatwave (with intensity of initial beams over
$10^{16}$~W/cm$^2$) in either centimeter-scale rarefied
($n_0\sim10^{18}$~cm$^{-3}$) or millimeter-scale dense plasmas
($n_0\gtrsim10^{19}$~cm$^{-3}$). We show that the plasma wave
driven {\it below} the plasma resonance, $\Omega<\omega_p$, chirps
the laser frequency in a very special way: near the center of each
laser beatnote the red-shifted sidebands are advanced in time with
respect to the blue-shifted ones. The GVD can compress thus
chirped beatnotes to a few-laser-cycle duration provided the laser
bandwidth tends to $\omega_0$. The effect of GVD is controllable:
proper adjustment of the plasma and laser parameters can reduce it
while preserving the desirable bandwidth. In this case, the
cascade compression can be made in two stages~\cite{KS} (see
Fig.~\ref{Fig1}): (i) a low-density plasma (the Modulator) with
$\Omega < \omega_{p(M)}$ is used for the FM of initially
two-frequency laser, and (ii) a higher-density plasma (the
Compressor) with $\omega_{p(C)} \gg \Omega$ serves for the
beatnote compression. Therefore, a train of sharp electromagnetic
spikes of intensity by several orders of magnitude higher than
ionization threshold for any medium can be generated. Similar
concept of using Raman cascades for radiation beams compression in
molecular gases has been successfully tested in
experiments~\cite{sokolov_jopb03} at low laser intensities
($\ll10^{14}$~W/cm$^2$). The technique of the EMC compression in
gases is not appropriate for applications requiring high laser
intensity. One such application is using multiple short laser
pulses with a tunable time delay for the coherent generation of
plasma waves~\cite{Umstadter,Dalla,Bonnaud}. Making a sequence of
several {\it independent} ultrashort high-intensity laser pulses
with the periodicity of less than one picosecond could be a major
experimental challenge~\cite{Umstadter}. The approach discussed in
our paper suggests a viable path to creating such pulse trains at
weakly relativistic intensity.

The outline of the paper is as follows. In Section~\ref{Sec2} we
derive the basic theoretical model (subsection~\ref{Sec2.1}) and
analyze the basic scalings for the EMC and the cascade compression
(subsection~\ref{Sec2.2}). In a realistic plasma, the EMC is a
complicated interplay between the sideband coupling through the
driven electron density perturbations, GVD of radiation,
nonlinearities due to the relativistic increase of an electron
mass, and SFRS. Fully nonlinear simulations presented in
Sec.~\ref{Sec3} account for all these effects and describe the
cascade development in either two-stage (subsection~\ref{Sec3.1})
or single-stage (subsection~\ref{Sec3.2}) compressor. Because the
longest time scale of the problem is only a few ion plasma
periods, parametric decay of the EPW~\cite{Mora} and consequent
plasma heating~\cite{Kaufman} are insignificant and thus ignored.
The simulation parameters of subsections~\ref{Sec3.1}
and~\ref{Sec3.2} are optimized so as to make the relativistic
nonlinearities and SFRS almost negligible. When the parameters of
the setup are not optimized, the SFRS can be seeded by the plasma
wake driven by the beatwave pulse of finite duration. Contribution
from the SFRS into the cascading process is discussed in the
subsection~\ref{Sec3.3}. In the subsection~\ref{Sec3.4} we show
how the beatnote compression helps in the EPW excitation via
relativistic bi-stability~\cite{bistability} with the initially
sub-threshold laser intensity. Conclusion gives the summary of the
results. In Appendix~\ref{AppendixA} the amplitude of the plasma
wake excited by a given detuned beatwave pulse is evaluated.

\section{ One-dimensional theory of EMC\label{Sec2}}
\subsection{\label{Sec2.1}Basic equations}

We assume that the laser duration does not exceed a few ion plasma
periods, so the ions are immobile and form a positive neutralizing
background. In one spatial dimension and in the limit of weakly
relativistic electron motion, Maxwell's equations and hydrodynamic
equations of electron fluid give the coupled equations for the
longitudinal and transverse momentum of
electrons~\cite{Gorbunov1}:
\begin{subequations}
\label{A1}
\begin{eqnarray}
\nonumber\lefteqn{\left(\frac{\partial^2}{\partial
t^2}+\omega_p^2\right)q_z = \frac{\omega_p^2}{2}q_zq^2
-\frac{c}{2}\frac{\partial^2q^2}{\partial z\partial t}}\\
&&{} -cq_z\left(1-\frac{q^2}{2}\right)\frac{\partial}{\partial
z}\left(\frac{\partial q_z}{\partial t}
+\frac{c}{2}\frac{\partial q^2}{\partial z}\right),\label{A1.1}\\
\nonumber\lefteqn{\left(\frac{\partial^2}{\partial
t^2}-c^2\frac{\partial^2}{\partial z^2}+\omega_p^2\right){\bf a} =
\frac{\omega_p^2}{2}q^2{\bf a}}\\
&&{} -c{\bf a}\left(1-\frac{q^2}{2}\right)\frac{\partial}{\partial
z}\left(\frac{\partial q_z}{\partial t} +\frac{c}{2}\frac{\partial
q^2}{\partial z}\right)\label{A1.2}.
\end{eqnarray}
\end{subequations}
Here, ${\bf a} \equiv {{\bf p}_e}_\perp/(m_ec)$ and $q_z\equiv{
p_e}_z /( m_ec)$ are the normalized transverse and longitudinal
components of the electron momentum, and $q^2\equiv q_z^2+{\bf
a}_\perp^2<1$. We take ${\bf a}\equiv \mathrm{Re}\,({\bf e}_0a)$,
where ${\bf e}_0=({\bf e}_x+i{\bf e}_y)/\sqrt2$ is a unit vector
of circular polarization; hence, $q^2=q_z^2+|a|^2/2$. In the 1D
approximation, conservation of the transverse canonical momentum
expresses the normalized momentum through the laser vector
potential, ${\bf A}_\perp=(m_ec^2/e){\bf a}$. The normalized
electron density perturbation,
\begin{equation}
\frac{n_e-n_0}{n_0}\equiv\frac{\delta
n}{n_0}\approx\frac{c}{\omega_p^2}\frac{\partial}{\partial
z}\left[\frac{\partial q_z}{\partial t} +
\frac{c}{2}\frac{\partial q^2}{\partial z}\right],\label{6}
\end{equation}
obeys the equation
\begin{eqnarray}
&&\left(\frac{\partial^2}{\partial
t^2}+\omega_p^2\right)\frac{\delta
n}{n_0}-\frac{c}{2}\frac{\partial^2q_zq^2}{\partial z\partial t} -
\frac{c^2}{2}\frac{\partial^2q_z^2}{\partial
z^2}\nonumber\\
&& {} +c\frac{\partial^2}{\partial z \partial
t}q_z\left(1-\frac{q^2}{2}\right)\frac{\delta
n}{n_0}=\frac{c^2}{4}\frac{\partial^2|a|^2}{\partial
z^2}\label{A2}
\end{eqnarray}
obtained through differentiating Eq.~(\ref{A1.1}) with respect to
$z$ and $t$. At the plasma entrance $z=0$, the amplitude of a
planar two-frequency laser beam is given by
\begin{equation}
\label{6.1}a(0,t)  =  e^{-i\omega_0t}\bigl[a_0(0,t) +
a_1(0,t)e^{-i\Omega t}\bigr],
\end{equation}
where $\Omega\approx\omega_p\ll\omega_0$. The ponderomotive
beatwave [the right-hand side (RHS) of Eq.~(\ref{A2})] produces an
electron density grating co-moving with the laser beams. The
moving index grating produces the cascade of laser sidebands,
\begin{equation}
a(z,t)  = e^{-i\omega_0t+ik_0z}\sum_{n=-\infty}^{+\infty}
a_n(z,t)e^{-in\Omega t+ink_\Omega z} ,\label{1}
\end{equation}
where $k_\Omega=\Omega/v_g$, and $v_g=k_0c^2/\omega_0$ is the
group velocity associated with the laser fundamental frequency.
$v_g$ is found from $d\equiv n_0/n_c=1-(v_g/c)^2$, where
$n_c=m_e\omega_0^2/(4\pi e^2)$ is a critical plasma density. The
amplitudes $a_n$ vary slowly in time and space on scales
$\Omega^{-1}$ and $k_\Omega^{-1}$.

To describe the nonlinear evolution of the cascade~(\ref{1}), we
take into account Eq.~(\ref{6}) and rewrite Eq.~(\ref{A1.2}) as
\begin{equation}
\label{7} \left(\frac{\partial^2}{\partial
t^2}-c^2\frac{\partial^2}{\partial z^2}\right)a +
\omega_p^2a\left(1-\frac{q^2}{2}\right)\left(1+\frac{\delta
n}{n_0}\right)=0.
\end{equation}
We retain in Eq.~(\ref{7}) the terms of order not higher than
$a^3$. Having in mind that $q_z\approx\delta n/n_0$, and that the
relativistic saturation of the beatwave-driven
EPW~\cite{rosenbluth_liu} occurs at $q_z\sim a^{2/3}$, we keep in
Eq.~(\ref{7}) the nonlinear terms of order $a^3$, $aq_z$,
$aq_z^2$, $aq_z^3$ and finally arrive at
\begin{equation}
\left(\frac{\partial^2}{\partial
t^2}-c^2\frac{\partial^2}{\partial
z^2}+\omega_p^2\right)a\approx\omega_p^2(R^a+R^q-C).\label{A4}
\end{equation}
The terms
\begin{subequations}
\label{A3}
\begin{eqnarray}
R^a & = & a|a/2|^2,\label{A3.1}\\
R^q & \approx & (a/2)(\delta n/n_0)^2 (1+\delta n/n_0)\label{A3.2}
\end{eqnarray}
\end{subequations}
originating from the relativistic mass correction of an electron
oscillating in the transverse ($R^a$) and longitudinal ($R^q$)
fields describe the relativistic self-phase-modulation of laser.
The leading nonlinear current term
\begin{equation} C  =
 a(\delta n/n_0) \label{A3.3}
\end{equation}
is responsible for the EMC and stimulated forward Raman
cascade~\cite{Skoric}.  Our earlier work~\cite{KS} assumed the
non-resonant EPW excitation ($\delta n/n_0\sim a^2$), so the
term~(\ref{A3.2}) was neglected. By including this term we include
the regime with relativistic saturation of the resonantly driven
EPW ($\delta n/n_0\sim a^{2/3}\gg a^2$).

We substitute the expansion~(\ref{1}) into Eq.~(\ref{A4}), replace
the variables $(z,t)$ by $(z,\xi)$ (where $\xi/v_g=t-z/v_g$ is a
retarded time, and $z$ is the propagation distance through
plasma), and collect the equal frequency terms. The resulting set
of coupled envelope equations
\begin{equation}
\left[2i\frac{\omega_n}{v_g}\frac{\partial}{\partial z}  -
\frac{d}{v_g^2}(\omega_n  -  \omega_0)^2\right] a_n  \approx
 k_p^2(C_n-R^a_n-R^q_n),\label{A6}
\end{equation}
where $k_p=\omega_p/c$, accounts for the propagation of sidebands
through plasma [the first term in the left-hand side (LHS)], the
GVD of the sidebands (the second term in the LHS), and the
sideband coupling through the nonlinearities (the RHS). To
evaluate the RHS of Eq.~(\ref{A6}) we have to specify the
nonlinear plasma response. The nonlinear electron density
perturbation is driven by the ponderomotive force [the RHS of
Eq.~(\ref{A2})] approximated as
$-(c/2)^2\sum_{l}(lk_\Omega)^2\rho_l(z,\xi)e^{ilk_\Omega\xi}$.
Here,
\begin{equation} \label{A5}
\rho_l=\sum_{m} a_ma^*_{m+l},
\end{equation}
and $|\partial\rho_l/\partial z|\ll k_\Omega|\rho_l|$. We expand
the density perturbation in the ponderomotive force harmonics,
\begin{equation}
\label{5} \delta n(z,\xi)=\frac12\sum_{l}\delta
n_l(z,\xi)e^{ilk_\Omega\xi},
\end{equation}
where $\delta n_{-l}=\delta n^*_l$, $|\delta n_l|\ll n_0$, and
$|\partial \delta n_l/\partial\xi|\ll k_\Omega|\delta n_l|$, and
assume that each density harmonic is driven by the corresponding
harmonic of the ponderomotive force. In the expansion~(\ref{5}),
the terms with $l=\pm1$ are the closest in frequency to the
natural modes of plasma oscillations. They produce the dominant
contribution to the cascade dynamics. Keeping in Eq.~(\ref{A2})
the terms of order not higher than $a^4$, and having in mind the
scaling $|\delta n/n_0|\sim a^{2/3}$ that holds in the case of
relativistic saturation of the resonantly driven
EPW~\cite{rosenbluth_liu}, we find that the amplitude
$N_e\equiv\delta n_{-1}(z,\xi)/n_c$ obeys the nonlinear equation
\begin{equation}
\left(\frac{i}{k_\Omega}\frac{\partial}{\partial \xi}
+\frac{\delta\omega}{\Omega}\right)N_e+R  =
\frac{d}{4}\rho_{-1}.\label{4}
\end{equation}
Here, $ \delta \omega = (\Omega^2-\omega_p^2)/(2\Omega)$ is the
beatwave detuning from the plasma resonance, and $R$ is
proportional to the nonlinear frequency shift due to the
relativistic mass increase of an electron oscillating in the
longitudinal and transverse electric fields,
\begin{equation}
\label{A7}
R=\frac{3}{16}N_e\left|\frac{N_e}{d}\right|^2+\frac18(\rho_0N_e+\rho_{-2}N_e^*).
\end{equation}
The initial condition for Eq.~(\ref{4}) is $N_e(z,-\infty)\equiv0$
(unperturbed plasma ahead of the pulse). Amplitudes of the
non-resonantly driven density harmonics ($l\not=\pm1$) are found
from the linearized Eq.~(\ref{A2}):
\begin{equation}
\label{A5new}\frac{\delta n_l(z,\xi)}{n_0} \approx
\frac{1}{2}\frac{
(\omega_l-\omega_0)^2}{(\omega_l-\omega_0)^2-\omega_p^2}\rho_l(\xi,z).
\end{equation}

Using equations~(\ref{A5new}), we evaluate the nonlinear terms in
the cascade equations~(\ref{A6}). We extract from the terms $C$ a
contribution from the EPW harmonics of orders $l\not=\pm1$ and
include it into the terms $R^a$. Therefore, only the contribution
from the near-resonant EPW harmonic $N_e$ determines the form of
the ``cascade'' nonlinearity $C_n$. And, only near-resonant EPW
harmonic is taken into account for evaluating the terms $R^q_n$.
The result is
\begin{subequations}
\label{A2.1}
\begin{eqnarray}
C_n & = & \frac{N_ea_{n-1}  + N_e^*a_{n+1}}{2d},\label{2.2}\\
R^a_n & \approx & \frac14 \sum_l a_{l+n}\rho_l \nonumber\\
& & {} -\frac14 {\sum_l}^\prime
a_{l+n}\rho_l\frac{(\omega_l-\omega_0)^2}{(\omega_l-\omega_0)^2 - \omega_p^2},\label{2.1}\\
R^q_n & \approx &
\frac{1}{4}\Biggl[\left|\frac{N_e}{d}\right|^2\left(a_n +
\frac{3}{4}\frac{N_e}{d}a_{n-1}+
\frac{3}{4}\frac{N_e^*}{d}a_{n+1}\right) \nonumber\\
& & {} + \frac12\left(\frac{N_e}{d}\right)^2
\left(a_{n-2}+\frac12\frac{N_e}{d}a_{n-3}\right)
\nonumber\\
& & {} + \frac12\left(\frac{N_e^*}{d}\right)^2
\left(a_{n+2}+\frac12\frac{N_e^*}{d}a_{n+3}\right)\Biggr]
\label{A8}.
\end{eqnarray}
\end{subequations}
The second term in the RHS of Eq.~(\ref{2.1}) comes from the
nonresonant EPW harmonics~(\ref{A5new}), prime means that the
terms with $l=\pm1$ are not included in the sum. The physical
meaning of the nonlinearities~(\ref{A2.1}) is as follows.
\begin{itemize}
\item The terms $C_n$ couple the neighboring laser sidebands
through the {\it near-resonantly-driven} harmonic of the EPW.
These terms describe the electromagnetic cascading and the
stimulated forward Raman cascade. \item The terms $R^a_n$ describe
the nonlinear frequency  shifts produced by the relativistic mass
increase of electron oscillating in the transverse fields and by
the {\it nonresonantly-driven} harmonics of EPW. \item The terms
$R^q_n$ describe the nonlinear frequency  shifts produced by the
relativistic mass increase of electron oscillating in the
longitudinal electric field of the {\it near-resonantly-driven}
harmonic of the EPW. Only when the EPW is driven resonantly and
reaches the relativistic saturation the term $R^q_n$ can dominate
$R^a_n$. In all the simulations that will follow in this paper
$R^q_n$'s are negligibly small.
\end{itemize}
Assuming $v_g\approx c$, we rewrite the set~(\ref{A6}) as
\begin{equation}
\left[\frac{2i}{k_0}\frac{\partial }{\partial z}    -
  d\left(\frac{\omega_n  -
\omega_0}{\omega_0}\right)^2 \frac{\omega_0}{\omega_n}\right] a_n
\! \approx \! d \frac{C_n \! - \! R^a_n \! - \!
R^q_n}{\omega_n/\omega_0}.\label{2}
\end{equation}
The boundary condition for Eqs.~(\ref{2}) is given by
Eq.~(\ref{6.1}). Equations~(\ref{4}) and~(\ref{2}) form the basis
of 1D spatio-temporal weakly relativistic theory of the
near-resonant plasma beatwave excitation. This theoretical model
encompasses the phenomena of the nonlinear excitation and
relativistic bi-stability of the EPW~\cite{bistability,Royal},
periodic FM of the laser~\cite{KS,sokolov_jopb03}, GVD of the
cascade components, and SFRS~\cite{Mori,Sakharov,Skoric}.

By the judicious choice of parameters the terms $C_n$ can be made
dominating in the RHS of Eq.~(\ref{2}), and laser spectral
broadening will occur exclusively due to the EMC. Despite the
large bandwidth of laser achieved in certain regimes of EMC, the
effect of GVD can be negligible (see the discussion at the end of
subsection~\ref{Sec2.2}). The laser amplitude can be then modified
in a separate, denser, plasma with the high GVD (the Compressor).
In this two-stage scenario, nonlinearities of the Modulator affect
primarily the laser phase, while in the Compressor the GVD
modulates the amplitude. Compression of the laser beatnotes in
plasma can result in the laser intensity so high as to give
$|a|^2\sim1$.

The laser sidebands in Compressor remain coupled through the
nonlinear frequency shifts, and the laser frequency bandwidth
keeps growing. Thus, the relativistic nonlinearities of plasma can
compete with the linear compression process. The Compressor
density is chosen so as to entirely exclude the possibility of
resonant plasma response: $\omega_{p(C)}$ is never close to an
integer multiple of $\Omega$. Provided $|n\rho_n|\ll 1$, the
amplitude of density perturbation at the $n^{\mathrm{th}}$
beatwave harmonic is described by Eq.~(\ref{A5new}) with $n_0$ and
$\omega_p$ replaced by $n_{0(C)}$ and $\omega_{p(C)}$. As the
electron density perturbations in the Compressor are nonresonant
and thus small, we neglect the terms $R^q_n$. Moreover, the terms
$C_n$ are absorbed by $R^a_n$ [that is, summation in the second
term of Eq.~(\ref{2.1}) is extended to $l=\pm1$, and $C_n$'s do
not show up in the Compressor equations]. We redefine the retarded
time as $\zeta/v_{g(C)}=t-z/v_{g(C)}$ (where $v_{g(C)}$ is the
group velocity of the laser fundamental component in the
Compressor plasma of density $n_{0(C)}\gg n_0$) and find that the
compression process can be described in terms of the coupled
nonlinear equations similar to Eqs.~(\ref{2}):
\begin{equation}
\Biggl[\frac{2i}{k_0}\frac{\partial }{\partial z} \! - \!
d_C\left(\frac{\omega_n  - \omega_0}{\omega_0}\right)^2
\frac{\omega_0}{\omega_n}\Biggr] a_n \! \approx \! -
d_C\frac{\omega_0}{\omega_n}R_{n}^{a(C)}. \label{2a}
\end{equation}
Here, $d_C=n_{0(C)}/n_c\ll1$ is the normalized Compressor density.
The boundary conditions are given by the solution of
Eqs.~(\ref{2}) at the Modulator exit, $a_n(z=z_{\cal M},\xi)$.

\subsection{Basic scalings for laser frequency modulation
and compression \label{Sec2.2}}

In the ideal two-stage compressor, the processes of EMC and
compression are separated.  The EMC develops in the Modulator
plasma with zero GVD, while in the dense Compressor plasma with
all the nonlinearities neglected the GVD compresses the radiation
beatnotes. It is instructive to derive the basic scalings for each
process because these approximate scalings will help to select the
optimal parameters of fully nonlinear simulations.

When both $R^a_n$ and $R^q_n$ are taken to be zero, and the GVD is
neglected ($d=0$) in Eq.~(\ref{2}), scaling laws for the EMC are
particularly simple. Assuming $\omega_n\approx\omega_0$, we derive
from Eqs.~(\ref{2}) a set of conservation laws: $\partial
\rho_l/\partial z = 0$. Hence, $\rho_l\equiv0$ for $l\not=0,-1$,
$\rho_0(z,\xi) \equiv |a_0(0,\xi)|^2+|a_1(0,\xi)|^2$,
$\rho_{-1}(z,\xi) \equiv a_1(0,\xi)a^*_0(0,\xi)$,  and, in the
co-moving frame, $N_e$ is independent of $z$ despite the evolution
of the laser phase. Thus simplified Eqs.~(\ref{2}) have the
analytic solution~\cite{Salomaa}
\begin{equation}
 a_n(z,\xi) =\sum_{\sigma=0,1} a_\sigma(0,\xi)
 e^{i(n-\sigma)(\psi+\pi)}
 J_{n-\sigma}(2W),\label{13}
\end{equation}
satisfying the initial condition~(\ref{6}) [here, $J_n(x)$ are the
Bessel functions, and $\psi(z,\xi)$ and $W(z,\xi)$ are the phase
and absolute value of the generating function $ w(z,\xi) \equiv W
e^{i\psi} = i(k_0z/4) N_e(\xi)$]. Substituting Eq.~(\ref{13}) into
Eq.~(\ref{1}) yields the expression for a train of phase-modulated
beatnotes, \[a(z,\xi)=\sum_{n=0,1}a_n(0,\xi) \cos [k_n\xi +
\varphi(z,\xi)],\] where $\varphi(z,\xi)=(k_0z/2)
|N_e(\xi)|\sin(\psi - k_\Omega\xi)$. The physical meaning of this
result is that, without GVD, the laser undergoes frequency
modulation only. The magnitude of the plasma wave depends only on
the laser amplitude which remains unchanged. This is valid for any
pair of $a_0(0,\xi), a_1(0,\xi)$ and the corresponding $N_e(\xi)$.

\begin{figure}[t]
\includegraphics[scale=1]{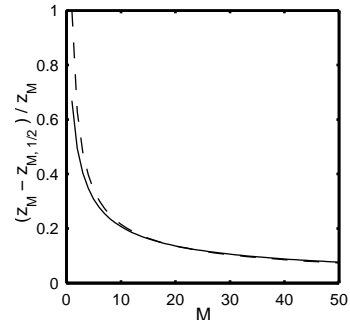}
\caption{\label{Fig3a}  Normalized distance between the points of
the first maximum and half-maximum of $J_M(z)$ (solid line) and
the scaling function $M^{-2/3}$ (dashed line). }
\end{figure}

The FM is periodic in time with the beat period $\tau_b$ when
$N_e(\xi)$ is almost constant [this is the case for
$|\partial\rho_{-1}/\partial\xi|\ll|(\delta\omega/c)\rho_{-1}|$].
To avoid the oscillations of $N_e$ with time due to the
relativistic dephasing~\cite{rosenbluth_liu,Tang,Gibbon2}, we take
$|\delta \omega|\gtrsim3(\omega_p/4)\sqrt[3]{3|\rho_{-1}|^2 /2}$.
Then, the term proportional to $\delta\omega$ dominates in the LHS
of Eq.~(\ref{4}), which yields $N_e(\xi)\approx
d[\rho_{-1}(\xi)/4](\Omega/\delta\omega)$; then, for real
$\rho_{-1}(\xi)$,
\begin{equation}
a \! = \! \sum_{n=0,1} \! a_n(0,\xi)  \cos \left[k_n\xi \! + \!
(k_0z/2)
  N_e(\xi)\cos( k_\Omega\xi)\right]. \label{16}
\end{equation}
From equation~(\ref{13}), a Modulator plasma slab of thickness
\begin{equation}
\label{8} z_{\cal M} \approx 2{\cal M}/(N_e k_0)
\end{equation}
produces ${\cal M}$ sidebands on either side of the fundamental,
and a frequency bandwidth $\Delta\omega\sim 2\sqrt{d}{\cal
M}\omega_0$. Conversely, ${\cal M}\sim r_e z_{\cal M}
\lambda_0|n_e-n_0|$, where $\lambda_0 = 2\pi c/\omega_0$ is the
fundamental laser wavelength, and $r_e=e^2/(m_ec^2)$ is the
classical electron radius.

As follows from Eq.~(\ref{16}), only for $\delta \omega < 0$ the
laser wavelength decreases with time near the amplitude maximum of
each beatnote (positive chirp). The GVD of plasma tends to
compress thus chirped beatnotes: the shorter (blue-shifted)
wavelengths catch up with the longer (red-shifted) wavelengths,
eventually building up the field amplitude near the center of each
beatnote. Thus, a sequence of sharp spikes is produced. If we
consider an unperturbed Compressor plasma of given density,
neglect the relativistic effects by setting
$R_{n}^{a(C)}\approx0$, and fix laser frequency and the number
$2{\cal M}$ of satellites, we find that the peak compression
occurs at a distance
\begin{equation}
\label{z_c}z_C\approx\frac{\pi/3}{k_0{\cal
M}}\left(\frac{\omega_0}{\omega_{p(C)}}\right)^2
\left(\frac{\omega_0}{\Omega}\vphantom{\frac{\omega_0}{\omega_{p(C)}}}\right)^2.
\end{equation}
This estimate assumes that the outer sidebands were initially
separated in time by roughly $\tau_b/2$ within one beatnote. To
catch up with the red sidebands at the beatnote center, the blue
sidebands need the propagation time $z_C/c \approx (c/\Delta
v_g)(\tau_b/2)$, where the group velocity mismatch is $\Delta v_g
\approx 2{\cal M} \Omega (\partial v_g/
\partial \omega)_{\omega_0} \approx (3 {\cal M} \Omega/k_0)(\omega_p/\omega_0)^2$.

The nonzero GVD of radiation in the Modulator plasma must be
properly accounted for. The cascade components can be
redistributed in time and space thus reducing coherence of the EPW
excitation and affecting the frequency chirp. Naively, the GVD can
become significant in the Modulator whose length $z_{\cal M}$ is
close to the compression length estimated from Eq.~(\ref{z_c})
with $\omega_{p(C)}\equiv\omega_p$. However, the higher-order
Stokes-anti-Stokes sidebands are generated later in plasma and
have less time to catch up with the fundamental. Recalling that
$|a_{\cal M}|\propto|J_{\cal M}(z)|$ in the Modulator, we define
the half-growth length $z_{{\cal M},1/2}$ at which $|a_{\cal M}|$
reaches a half of its maximum value, $|J_{\cal M}(z_{{\cal
M},1/2})|=|J_{\cal M}(z_{\cal M})|/2$. Thereby, compression
effectively takes place over the shorter distance $\Delta z\approx
z_{\cal M} - z_{{\cal M},1/2} < z_{\cal M}$. The analytic formula
$\Delta z\approx {\cal M}^{-2/3}z_{\cal M}$ accurately fits
$\Delta z$ for ${\cal M} \gtrsim 5$ (see Fig.~\ref{Fig3a}).
Therefore, $\Delta z \ll z_{\cal M}$ for ${\cal M} \gg 1$. Hence,
for
\begin{equation}
\label{GVD_suppress} z_{\cal M} \ll {\cal
M}^{2/3}z_C\approx(\lambda_0/6){\cal M}^{-1/3}d^{-2},
\end{equation}
the effect of the GVD is negligible in the Modulator because the
distance $\Delta z$ actually available for the compression is less
than $z_C$. Otherwise, if $z_{\cal M}\gtrsim {\cal M}^{2/3}z_C$,
the GVD in the Modulator becomes important.
\begin{figure}[t]
\includegraphics[scale=1]{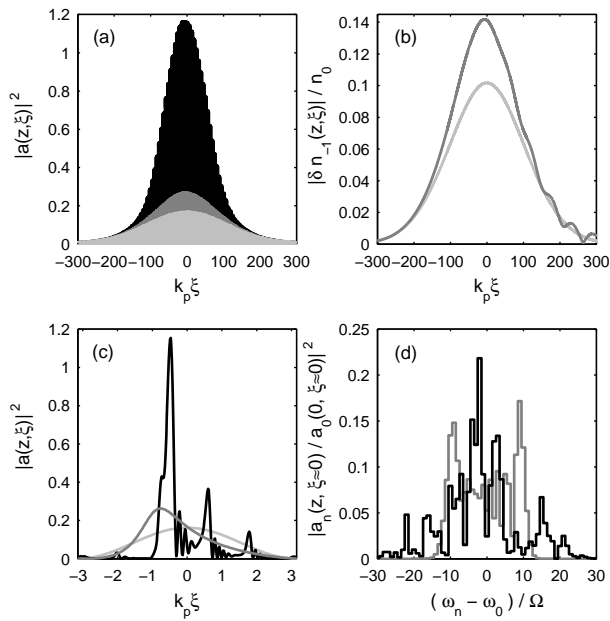}
\caption{\label{Fig2} The two-stage cascade compression. The
physical quantities are shown at the entrance ($z=0$, light gray)
and exit of the Modulator ($z=z_8$, medium gray), and after the
Compressor ($z=z_8+z_C$, black). (a) The laser pulse intensity
(the time window contains about 100 beatnotes). (b) The normalized
amplitude of the near-resonant EPW, $\delta n_{-1}/n_0=N_e/d$. (c)
The beatnote intensity near the laser pulse center; one beat
period near $\xi=0$ is shown. (d) The laser spectra near $\xi=0$.
The nonlinearities and GVD in both plasmas are included. }
\end{figure}

Another manifestation of the GVD in the Modulator is the SFRS
seeded due to finite duration of the beatwave pulse. The
stimulated forward Raman cascading~\cite{Skoric} can interfere
with the process of phase modulation and contaminate the laser
frequency chirp. Reduction in the compression efficiency can
follow. The effect of SFRS is examined  in
subsection~\ref{Sec3.3}.

\section{Nonlinear simulations of the EMC\label{Sec3}}

\subsection{The two-stage cascade compressor\label{Sec3.1}}

We model the EMC by numerically solving the set of coupled
nonlinear equations~(\ref{4}) and (\ref{2}) with the boundary
condition
\begin{equation}
\label{2b} a_0(0,\xi)=a_1(0,\xi)=Ae^{-\xi^2/(c\tau_L)^2}
\end{equation}
for the laser sidebands, and $N_e(z,\xi=-\infty)\equiv0$ for the
EPW. The beatnote compression in the second stage is modeled by
numerically solving the Compressor equations~(\ref{2a}).  All the
nonlinearities associated with the effects of relativistic mass
correction and non-resonant electron density perturbations are
retained in the modeling of both stages. In all the simulations
below, the fundamental laser wavelength is fixed at
$\lambda_0=0.8$~$\mu$m.

The two-stage compression starts with the initial laser amplitude
$A = 0.2$, the Modulator density $n_0=8.75\times10^{17}$~cm$^{-3}$
(hence, $d=5\times10^{-4}$), and $\delta\omega = - 0.1 \omega_p$.
The laser pulse duration is $\tau_L=4.5$~ps (about half the ion
plasma period for a fully ionized Helium). Having chosen the
maximum density perturbation
$|N_e(z=0)|_{\max}\approx0.5\times10^{-4}$ and resulting spectral
width of the laser (${\cal M}\approx8$ sidebands on each side), we
find the Modulator length $z_8\approx4.1$~cm (such interaction
length could be implemented in a plasma
channel~\cite{EsareyIEEE}).

The simulation results are shown in Fig.~\ref{Fig2}. From the plot
(a) it is seen that the peak laser intensity at the Compressor
exit ($z=z_8+z_C$) is by a factor of 7.2 larger than at the
Modulator entrance ($z=0$). The increase in intensity results from
the shown in plot (c) beatnote compression from the initial
duration of $\tau_{b(in)}\approx120$~fs to
$\tau_{b(out)}\approx13$~fs (roughly 5 laser cycles). Compressor
plasma has the density $n_{0(C)}=25n_0$ and length $\approx0.0275
z_8\approx 1.1$~mm (such a short dense plasma can be created by
ablation of a microcapillary~\cite{Suckewer}).

The inequality~(\ref{GVD_suppress}) is very well satisfied for the
Modulator parameters. Consequently, the beatnote pre-compression
seen in the plot~(c) is quite insignificant. Plot~(b) shows that
$N_e(z,\xi)$ also reveals almost negligible variation with $z$ in
the Modulator. Thereby, according to the plots~(b) and~(c), the
EMC develops in accordance with the scenario outlined in
subsection~\ref{Sec2.2}.

Compression in the second stage clearly proceeds in the nonlinear
regime. The laser amplitude becomes relativistic ($|a|\to1$), and
the nonlinear frequency shifts in Eqs.~(\ref{2a}) couple the laser
sidebands and further increase the laser bandwidth.
Figure~\ref{Fig2}(d) shows that the resulting frequency spectrum
is at least twice as broad if compared with that at the Modulator
exit. As a consequence, the linear formula~(\ref{z_c})
overestimates $z_C$ by a factor of three since it ignores both
pre-compression of the pulse in the Modulator and additional
bandwidth increase in the Compressor. Also, quality of the
compressed beatnotes is not perfect: instead of a single sharp
spike, one can observe a multi-spike structure in
Fig.~\ref{Fig2}(c), the distance between the spikes being roughly
$\tau_b/5$. One can relate this structure to the phase modulation
occurring due to the electron density perturbation at fifth
harmonic of the beatwave frequency, which is the closest to the
natural mode of the Compressor plasma oscillations. Due to this
effect, one beatnote is not gradually compressed into one spike
but rather splits into five spikes, of which the one located near
the original beatnote maximum has the largest amplitude.

A number of issues are yet to be addressed before the theory of
the cascade compression is complete. The neglected effects of
transverse evolution of the laser, such as relativistic
self-focusing and cascade focusing~\cite{Gibbon2}, are dominant at
high plasma density in the Compressor and are potentially adverse.
But, we find the 1D scenario of the two-stage compression
conceptually simple and useful for understanding the underlying
phenomena. In the next subsection we consider the single-stage
approach which assumes concurrent cascading and compression in the
same low-density plasma.

\subsection{The single-stage cascade compressor\label{Sec3.2}}

Increasing plasma density in the Modulator increases the GVD.
Therefore, we can explore the idea of compressing the beatnotes
concurrently with generating the sidebands. In the following set
of simulations the electron density is doubled,
$n_0=1.75\times10^{18}$~cm$^{-3}$. Cascade compression is
simulated in plasma of the same length, $z_8=4.1$~cm, and with the
same initial density perturbation
$|N_e(z=0)|_{\max}\approx0.5\times10^{-4}$ as in the previous
subsection. The laser amplitude is $A\approx0.071$, and the
beatwave detuning is $\delta\omega = - 0.025 \omega_p$. As we
shall see in subsection~\ref{Sec3.3}, the SFRS manifestation can
be large in this regime. Appropriately low seed level can be
achieved with the beatwave pulse envelope varying slowly on the
time scale $\delta\omega^{-1}$. We choose the beatwave pulse
$\tau_L=14.25$~ps long (which is about three ion plasma periods
for a fully ionized Helium), which corresponds to
$|\delta\omega\tau_L|\approx27$. The given initial intensity on
axis and laser duration require the pulse energy of about 5~J in
the focal spot of 30~$\mu$m radius.
\begin{figure}[t]
\includegraphics[scale=1]{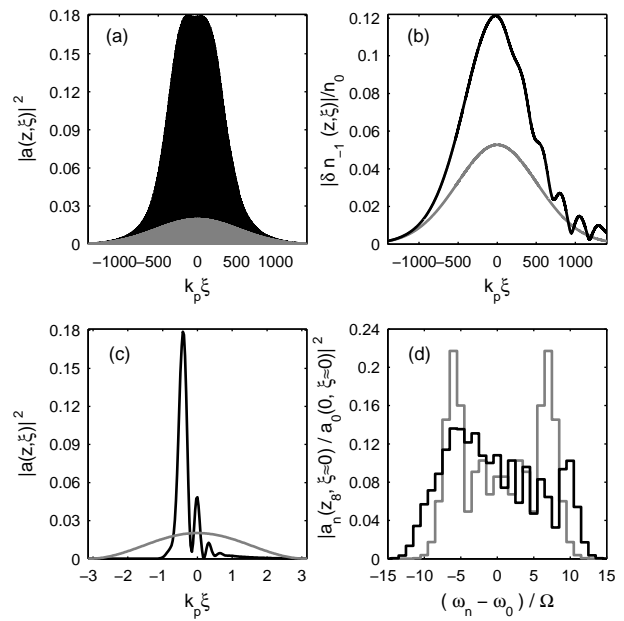}
\caption{\label{Fig4} The single-stage compressor with concurrent
EMC and compression. In plots (a) - (c), the physical quantities
are shown at the entrance ($z=0$, gray) and exit of the plasma
($z=z_8$, black). (a) The laser pulse intensity (the time window
contains about 500 beatnotes). (b) The normalized amplitude of the
near-resonant EPW, $\delta n_{-1}/n_0=N_e/d$. (c) The beatnote
intensity near the laser pulse center, $\xi=0$. (d) The laser
spectra near $\xi=0$ with (black) and without (gray) GVD and all
nonlinear frequency shifts.}
\end{figure}

The linear estimate of the effective compression
length~(\ref{z_c}) shows that the inequality~(\ref{GVD_suppress})
almost breaks under the simulation parameters. Hence, a beatnote
compression is large at the plasma exit. Figure~\ref{Fig4}(a)
shows that the resulting peak intensity is by an order of
magnitude larger than at the plasma entrance. Initial duration of
the laser beatnote, as shown in Fig.~\ref{Fig4}(c), is reduced
from $\tau_{b(in)}\approx85$~fs by roughly a factor of 10 (to
roughly 3 laser cycles). The spectral features of the EMC are
different from those obtained with the nonlinear frequency shifts
and the GVD neglected [i.e., with $d=0$ in Eqs.~(\ref{2})]. A red
asymmetry of the cascade spectrum is seen in Fig.~\ref{Fig4}(d).
Some spectral broadening versus the case of $d=0$ can be
attributed to the self-phase-modulation produced by the nonlinear
frequency shifts. Figure~\ref{Fig4}(b) shows that the electron
density perturbation is not an integral of motion. Its final
amplitude is roughly twice the initial, and a small-amplitude wake
is left behind the train of compressed spikes which can be
recognized as a signature of SFRS. Despite of this, neither the
self-phase-modulation nor the SFRS are adverse for the cascade
compression.

The transverse evolution of the cascade is a matter of high
importance for the experimental verification of this compression
scheme. For instance, electron density perturbations can
significantly lower the nonlinear focusing threshold of
counter-propagating laser beams~\cite{ShvetsPukhov}. The
co-propagating cascade of electromagnetic beams also experiences
enhanced focusing in both 2D (planar) and 3D (cylindrical)
geometry if $\Omega<\omega_p$
(Refs.~\onlinecite{Gibbon1,Gibbon2}). We shall give here a few
necessary estimates (effects of transverse evolution will be given
a detailed consideration in the upcoming publications). When the
beatwave is downshifted by
$\delta\omega=-3(\omega_p/4)\sqrt[3]{3|\rho_{-1}|^2 /2}$, the
self-focusing threshold in the planar 2D
geometry~\cite{Gibbon2,Gibbon1} is $a_0(k_px_0/10)^3\ge0.064$,
where $x_0$ is a laser focal spot size ($a_0=a_0e^{-x^2/x_0^2}$).
So, the sub-threshold regime under the parameters of
Figs.~\ref{Fig2} and~\ref{Fig4} requires the spot size
$x_0<40$~$\mu$m. If we loosely translate $x_0$ into the radius of
the laser focal spot in the cylindrical geometry, we find that the
refraction-limited interaction length $2z_R=(2\pi/\lambda_0)x_0^2$
is less than 1.2 cm. However, the required $z_8\approx4.1$~cm can
be achieved by means of the plasma channel
guiding~\cite{EsareyIEEE}.

\begin{figure}[t]
\includegraphics[scale=1]{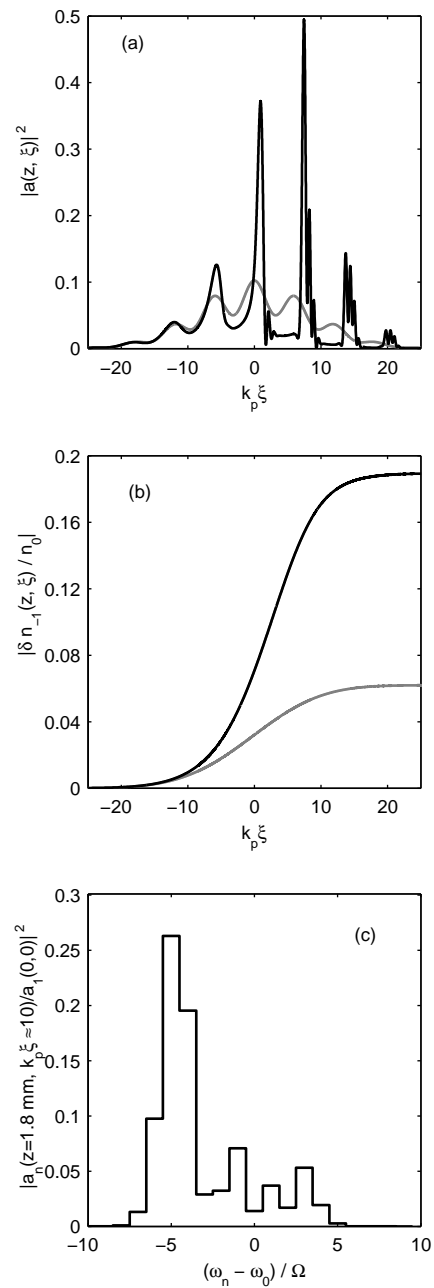}
\caption{\label{Grigsby} The ECM of the two-color short-pulse
(90~fs) laser~\cite{Grigsby} in a dense
($1.35\times10^{19}$~cm$^{-3}$) plasma. Physical quantities are
shown at the plasma entrance ($z=0$, gray lines) and exit
($z=1.8$~mm, black lines). }
\end{figure}

In the regime of cascade compression considered above, eliminating
potentially adverse effects of laser self-phase-modulation and
SFRS required complying with some hard restrictions on the laser
pulse amplitude, duration, and beatwave frequency detuning
($|a|^2\ll1$, and $|\delta\omega\tau_L|\ggg1$). The next set of
simulations shows that these conditions are desirable but not
necessary for the manifestation of the effect. The cascade
compression can be observed even for ultrashort ($\sim100$~fs)
beatwave pulses propagating in a dense plasma
($n_0\sim10^{19}$~cm$^{-3}$) where neither GVD nor relativistic
nonlinearities are small. We consider the evolution of a two-color
ultrashort laser~\cite{Grigsby} whose energy is initially
distributed between the fundamental (97\%) and the Stokes (3\%)
components. The laser frequencies are
$\omega_0=2.356\times10^{15}$~s$^{-1}$ ($\lambda_0=0.8$~$\mu$m)
and $\omega_1=2.159\times10^{15}$~s$^{-1}$
($\lambda_1=0.873$~$\mu$m). Assuming that
$\omega_0-\omega_1=0.95\omega_p$, we derive the plasma density
$n_0=1.35\times10^{19}$~cm$^{-3}$; hence, $d=7.725\times10^{-3}$.
We choose $a_0\approx0.3$, and $a_1\approx0.048$. At $z=0$, the
laser pulse is Gaussian~(\ref{2b}) with a duration
$\tau_L\approx90$~fs. In this case
$|\delta\omega\tau_L|\approx0.9$, and the beatwave pulse amplitude
is not slowly varying. Nevertheless, under these seemingly
unfavorable conditions, the EMC develops very effectively and the
intensity contrast of the amplitude-modulated laser pulse grows
rapidly. The plasma length is chosen so as to produce 5 sidebands
on either side; the plasma length is bounded from above by
$z_5\approx2.5$~mm, while the compression length evaluated from
formula~(\ref{z_c}) for ${\cal M}=5$ gives the lower bound,
$z\approx0.45$~mm. The most spectacular features of the EMC shown
in Fig.~\ref{Grigsby} are observed at $z\approx1.8$~mm. At the
plasma border, $z=0$, the intensity variation of the two-color
laser is about 50\%, while at $z\approx1.8$~mm very deep amplitude
modulation develops with the intensity contrast ratio reaching a
factor of 25. The mostly affected are the beatnotes in the tail of
the laser pulse; they are compressed to roughly a quarter of a
plasma period. The laser nonlinear evolution boosts the amplitude
of the plasma wake (which is increased by a factor of 3). The
laser spectrum broadens and reveals a red shift by about
$-\omega_0/2$. This is a clear indication of the forward
stimulated Raman cascade~\cite{Skoric} which, as appears in this
simulation, does not prevent the beatnote compression. Moreover,
the laser frequency red-shifts towards the pulse tail. The
red-shifted field components that form the compressed beatnotes in
the tail move slower than those in the pulse head. Thus, as seen
in Fig.~\ref{Grigsby}(a), the beatnotes in the tail accumulate a
considerable time delay (about a quarter of the beat period) with
respect to the initial positions of their maxima. This time delay
is in agreement with the frequency shift $-\omega_0/2$.

The simulations presented in this subsection demonstrate the
robustness of the EMC in the conditions when the GVD is large, and
the nonlinear processes of the relativistic self-phase modulation
and the SFRS interfere the cascading process. Slow variation of
the beatwave pulse envelope is therefore helpful but not necessary
for the cascade compression.

\subsection{Manifestation of
SFRS in cascade compressor \label{Sec3.3}}

Equations~(\ref{2}) admit the longitudinal transfer of
electromagnetic energy in the co-moving frame. Therefore, the
laser amplitude modulation may result not only from the EMC with
concurrent compression of beatnotes but also as a consequence of
the SFRS instability~\cite{Mori,Sakharov,Skoric} (also referred to
as the 1D resonant modulational instability~\cite{Andreev}). The
SFRS is different from the EMC. The latter is merely a phase
modulation which may proceed in the absence of GVD.

The SFRS is a resonant process seeded by the electron density
perturbations oscillating at the plasma frequency $\omega_p$. The
instability bandwidth is much narrower than $\omega_p$ even in the
case of relativistically strong pump, $a_0\sim1$
(Ref.~\cite{Sakharov}). And, in the examples of Figs.~\ref{Fig2}
and~\ref{Fig4}, the SFRS bandwidth is much lower than the absolute
value of the beatwave frequency detuning $\delta\omega$. However,
an electron plasma response to the laser beatwave {\it always}
contains a component oscillating at $\omega_p$. This component is
due to the finite duration of the beatwave pulse, and its
amplitude is governed by the product $|\delta\omega\tau_L|$. These
resonant density perturbations can be enhanced by the SFRS to a
level comparable to that of a non-resonant plasma response, and
can interfere the phase modulation process. Hence, the effect of
the SFRS is adverse and should be avoided by the judicious choice
of laser and plasma parameters.

The seed level for the SFRS can be reduced by taking
$|\delta\omega\tau_L|\ggg1$. For example, parameters used in
Figs.~\ref{Fig2} and~\ref{Fig4} correspond to
$|\delta\omega\tau_L|\sim30$ and reveal no SFRS manifestation: no
considerable plasma wake is left behind the laser at $z=z_8$.
Hence, the plasma response is almost entirely non-resonant in
these simulations.

\begin{figure}[t]
\includegraphics[scale=1]{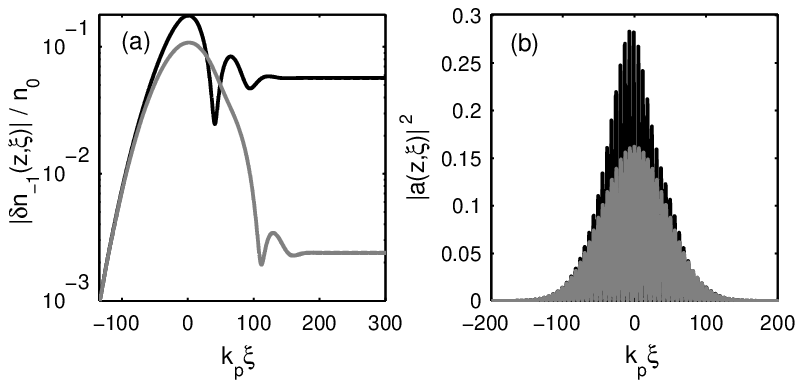}
\caption{\label{Fig2a} Electron density perturbation (a) and the
temporal profile of laser intensity (b) for the parameters the
same as of Fig.~\ref{Fig2} except the laser duration reduced by a
factor of 2.5, $\omega_p\tau_L\approx100$. Gray color --- $z=0$,
black --- $z=z_8$.}
\end{figure}

Under parameters of Fig.~\ref{Fig2}, reduction in the beatwave
pulse duration by a factor of 2.5 ($|\delta\omega\tau_L|=10$)
produces visible enhancement of the plasma wake that can be
attributed to the SFRS manifestation (see Fig.~\ref{Fig2a}). At
the plasma border, the wake amplitude is $\delta
n(z=0)\equiv\delta n_s\approx2.4\times10^{-3}n_0$. Taking $\delta
n_s$ as the SFRS seed amplitude, we can theoretically evaluate the
amplification factor by using the formula~(4.12) of
Ref.~\cite{Sakharov}, $\ln|\delta n(z)/\delta n_s|\approx
(2/c)\bigl(z\int_{-\infty}^{+\infty}\Gamma_0^2(\tau)\,d\tau\bigr)^{1/2}$.
This expression takes into account the laser temporal profile and
is valid for $z\gg\xi\gg c\tau_L$;
$\Gamma_0^2(\tau)=(1/8)(\omega_p^4/\omega_0^2)a_0^2(\tau)$ stands
for the instantaneous growth rate. The theoretical estimate of
amplification factor is $\ln|\delta n(z_8)/\delta
n_s|\approx3.36$. On the other hand, comparing the wake amplitudes
at the entrance ($z=0$) and at the exit ($z=z_8$) of the plasma
gives the amplification factor of $\ln|\delta n(z_8)/\delta
n_s|\approx 3.2$, which is very close to the analytical estimate.
We have found that the theory and simulation agree for
$80<\omega_p\tau_L<300$. Therefore, throughout this range, plasma
wakes are excited almost entirely by the SFRS. Remarkably, the
maximum laser intensity, as well as the shape of individual
beatnotes, is almost the same at $z=z_8$ for the parameters of
Figs.~\ref{Fig2} and~\ref{Fig2a}. Hence, in the considered
parameter range the effect of SFRS has a negligible effect on the
laser evolution.

\begin{figure}[t]
\includegraphics[scale=1]{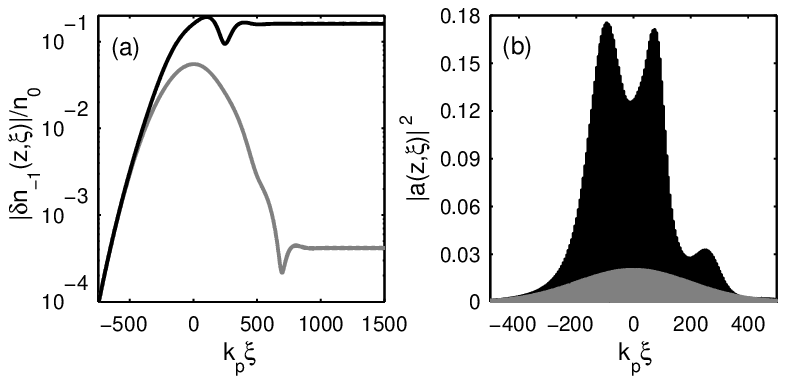}
\caption{\label{Fig4a} Electron density perturbation (a) and the
temporal profile of laser intensity (b) for the parameters the
same as of Fig.~\ref{Fig4} except the laser duration reduced to
$\omega_p\tau_L\approx440$, and $|\delta\omega\tau_L|=10$. Gray
color --- $z=0$, black --- $z=z_8$.}
\end{figure}

Oppositely to the just discussed case of rarefied plasma, reducing
the laser duration by the same factor 2.5 under the parameters of
Fig.~\ref{Fig4} (i.e., plasma twice as dense versus that of
Figs.~\ref{Fig2} and~\ref{Fig2a}) causes significant enhancement
of SFRS. Figure~\ref{Fig4a}(a) shows the plasma wake amplification
by a factor of $\ln|\delta n(z_8)/\delta n_s|\approx 6.05$.
Theoretical estimate of the SFRS gain gives 4.2; this discrepancy
can be partially explained by the laser amplitude growth due to
the beatnote compression. Figure~\ref{Fig4a}(b) demonstrates the
strong deformation of the beatwave intensity profile. The spectral
content of the electromagnetic cascade varies considerably and can
exhibit either overall red- or blue-shift at different $\xi$ in
the window $-100<k_\Omega\xi<100$.

In conclusion, to get rid of the SFRS, one should keep the SFRS
seed low by keeping the product $|\delta\omega\tau_L|$ large. As
the simulations show, it should be larger than 20; this may
require the beatwave pulse duration of several picoseconds or
larger.

\subsection{Relativistic bi-stability of EPW\label{Sec3.4}}

Cascade compression is a perfect tool for studying threshold
phenomena. One of them, the relativistic bi-stability (RB) of the
EPW driven by the long ($|\delta\omega\tau_L|\gg1$) beatwave pulse
with $\Omega<\omega_p$, is considered in this subsection. The RB
results in the excitation of large-amplitude plasma
wakes~\cite{bistability}. The intensity threshold should be met
for the RB to occur. The threshold is multi-faceted: it is
determined by the beatwave frequency detuning $\delta\omega$, the
laser amplitude, shape, duration, contribution from the plasma
wave harmonics {\it etc}. Various aspects of the RB in the
approximation of the given driver $\rho_{-1}(\xi)$ are addressed
in the forthcoming publication~\cite{Royal}.
\begin{figure}[h]
\includegraphics[scale=1]{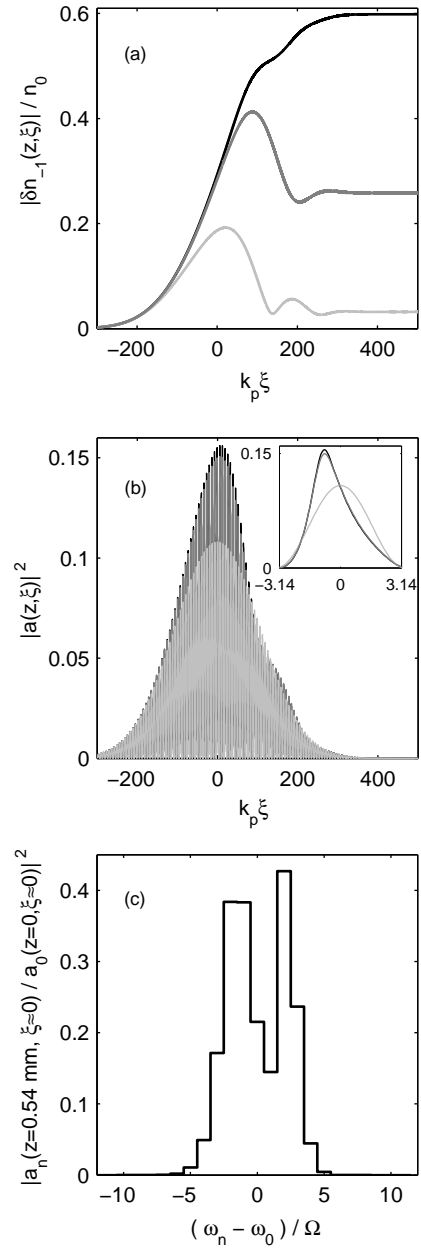}
\caption{\label{FigBSTB} Relativistic bi-stability of the EPW.
Magnitude of the near-resonant electron density perturbation [plot
(a)] and the laser intensity [plot (b)] are shown at $z=0$ (light
gray), $z=0.52$~mm (medium gray), and $z=0.54$~mm (black). Inset
in plot~(b): one beatnote selected near $\xi=0$. Plot~(c): the
laser spectrum at $z=0.54$~mm. The RB threshold is crossed at
$z\approx0.53$~mm. Crossing the threshold increases the EPW
amplitude by a factor of 2.3.}
\end{figure}

Evolution of the laser amplitude becomes important when a laser
propagates over a considerable distance in the plasma. Strong
distortion of the beatwave temporal profile within a finite
distance (few millimeters) in a dense ($n_0\sim10^{19}$~cm$^{-3}$)
plasma does have adverse consequences for the amplitude- and
phase-sensitive process of RB. On the other hand, we demonstrate
below that the beatnote compression helps to cross the RB
threshold in the case of initially sub-threshold laser amplitude.
We start the simulation with the parameters of the numerical
example from Ref.~\cite{bistability}: simulation starts at $z=0$
in a plasma with a density $n_0=10^{19}$~cm$^{-3}$, the beatwave
pulse having a Gaussian temporal profile~(\ref{2b}) with
$\omega_p\tau_L\approx212$ and $\delta\omega=-0.05\omega_p$, and
the laser fundamental wavelength being $\lambda_0=0.8$~$\mu$m
(then, $d\approx0.0057$). Solution of Eq.~(\ref{4}) with a given
driver (equivalent to the calculations of plasma response at the
entrance point $z=0$) show that the RB threshold is
$A^2\approx0.03375$ (16\% lower than in
Ref.~\onlinecite{bistability}). This threshold corresponds to the
normalized peak intensity $|a|^2_{\max}\approx0.135$. To
demonstrate how this threshold is crossed in the course of laser
evolution in plasma, we start the simulation with a sub-threshold
value of the laser intensity, $A^2\approx0.027$. The simulation
results are shown in Fig.~\ref{FigBSTB}.

As the laser travels through plasma, the cascade compression of
the beatnotes locally increases the intensity, and the RB
threshold is crossed at a distance $z\approx0.53$~mm. At this
point, the magnitude of the electron density perturbation jumps
abruptly by a factor of 2.3 [$\delta n_{-1}/n_0$ immediately
before and after crossing the RB threshold is shown in
Fig.~\ref{FigBSTB}(a)]. After that point, the resonant density
perturbation grows steadily to roughly 80\% of background density.
At $z\approx1.2$~mm the beatwave amplitude distortion becomes so
strong as to destroy the coherence of the plasma response, and
$\delta n_{-1}$ drops sharply. Importantly, at the point where the
RB threshold is met, the beatnote amplitude is not very different
from sinusoidal [see the inlay in Fig.~\ref{FigBSTB}(b)], and the
shape of the beatwave pulse is not much different from the initial
Gaussian. The normalized intensity at which the RB occurs in the
simulation is $|a|^2_{\max}\approx0.145$. The amplitude of the
beatwave pulse immediately before ($z=0.52$~mm) and after
($z=0.54$~mm) crossing the RB threshold is almost the same, as can
be seen in the plot~\ref{FigBSTB}(b). As follows from
Fig.~\ref{FigBSTB}(c), the beatnote compression necessary for
reaching the RB threshold is achieved at the laser bandwidth
roughly equal to $\omega_0/3$.

Therefore, the proposed model of EMC in plasmas with nonzero GVD
is able to demonstrate the effect of relativistic bi-stability in
the dynamic simulations with initially sub-threshold laser
amplitude.

\section{Conclusion}
In this paper, we have developed a nonlinear model that describes
the evolution of laser beatwave and electron density perturbations
in time and in 1D in space in the weakly relativistic regime.
Electromagnetic spectrum evolution and the effects of finite group
velocity dispersion are accurately modeled. The model includes the
nonlinear frequency shifts related to the relativistic corrections
of electron mass and the harmonics of the electron density
perturbations. It also takes into account the spatio-temporal
evolution of the near-resonantly driven electron density
perturbation. The theoretical model also describes a number of
nonlinear effects important for the implementation of plasma
beatwave accelerator. It is found that, for the beatwave
downshifted in frequency from the plasma resonance, the
electromagnetic cascading produced by the near-resonant electron
density perturbations leads to the compression of the laser
beatnotes, which finally transforms the beatwave pulse into a
train of sharp (few-laser-cycle) electromagnetic spikes separated
by the beat period in time and space. A train of electromagnetic
pulses useful for the particle acceleration
applications~\cite{Umstadter,Dalla,Bonnaud} can be
self-consistently created. We are also able to demonstrate how the
electron plasma wave of large amplitude can be excited due to the
effect of relativistic bi-stability even in the case of initially
sub-threshold beatwave pulse. The work is supported by the U.S.
Department of Energy under Contracts No. DE-FG02-04ER54763 and
DE-FG02-04ER41321, by the National Science Foundation grant
PHY-0114336 administered by the FOCUS Center at the University of
Michigan, Ann Arbor.

\appendix

\section{Electron density perturbation driven by a given beatwave pulse\label{AppendixA}}

\begin{figure}[t]
\includegraphics[scale=1]{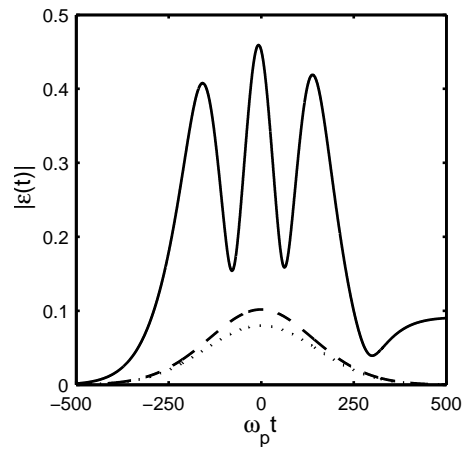}
\caption{\label{AppB} Temporal evolution of the EPW driven by the
two-frequency laser pulse of duration $\omega_p\tau_L=300$. The
laser intensity profile $|a_0(t)|^2+|a_1(t)|^2$ is shown with a
dotted line. Numerical solution to Eq.~(\ref{B1}) gives the EPW
amplitude in the resonant ($\Omega=\omega_p$, solid line) and
near-resonant ($\Omega=0.9\omega_p$, dashed line) cases.}
\end{figure}

In this Appendix we evaluate analytically and numerically the
initial density perturbations in the ranges of laser parameters
(amplitude, duration, and the beat frequency) relevant to the
numerical examples of this paper. We consider the excitation of
EPW at the plasma border $z=0$ and take the Gaussian temporal
profile of laser~\ref{2b}. Then, the normalized amplitude
$\varepsilon(t)=\delta n(t)/n_0$ of near-resonantly driven
electron density perturbation obeys the weakly nonlinear equation
\begin{equation}
[\Omega^{-1}\partial /\partial t - i\delta\omega(t)]\varepsilon=-i
(A/2)^2e^{-2t^2/\tau_L^2},\label{B1}
\end{equation}
where
$\delta\omega(t)=\delta\tilde{\omega}+(3/16)|\varepsilon(t)|^2+(A/2)^2e^{-2t^2/\tau_L^2}$,
$\delta\tilde{\omega}=\delta\omega/\Omega$,
$|\delta\omega(t)|\ll1$. We will solve this equation numerically
with the initial condition $\varepsilon(-\infty)=0$ (plasma is
quiescent before the laser pulse arrival).

Note, that Eq.~(\ref{B1}) can be solved analytically by means of
the perturbational approach when
$|\delta\tilde{\omega}_l|>\Delta\equiv(1/2)\sqrt[3]{3(A/2)^4}$
(Ref.~\cite{Tang}), and $\delta\omega\tau_L\ggg1$. Then,
$\varepsilon(t)$ adiabatically follows the temporal profile of
ponderomotive force, and, in the zero-order approximation, the
time derivative can be omitted in Eq.~(\ref{B1}). First-order
approximation is obtained via linearization of Eq.~(\ref{B1}) with
$|\varepsilon(t)|^2\approx\bigl[(A/2)^4/\delta\tilde{\omega}^2\bigr]e^{-4t^2/\tau_L^2}$.
The solution reads
\begin{eqnarray}
\nonumber \varepsilon(t) & \approx & -i(a/2)^2 e^{i\delta\omega t
-i\Omega\tau_L\Phi(t)}\\
 & \times &\int_{-\infty}^t
e^{-2\tau^2/\tau_L^2- i\delta\omega\tau+
i\Omega\tau_L\Phi(\tau)}\, d\tau,\label{B2}
\end{eqnarray}
where $\Phi(x)  = \kappa_1\mathrm{erfc}(2 x/\tau_L)+
\kappa_2\mathrm{erfc}(\sqrt2 x/\tau_L)$,
$\mathrm{erfc}(y)=(2/\sqrt{\pi})\int_y^{+\infty}e^{-t^2}dt$,
$\kappa_1=(3\sqrt{\pi}/64)(A/2)^4/\delta\tilde{\omega}^2$, and
$\kappa_2=\sqrt{\pi/8}(A/2)^2$.

\begin{figure}[t]
\includegraphics[scale=1]{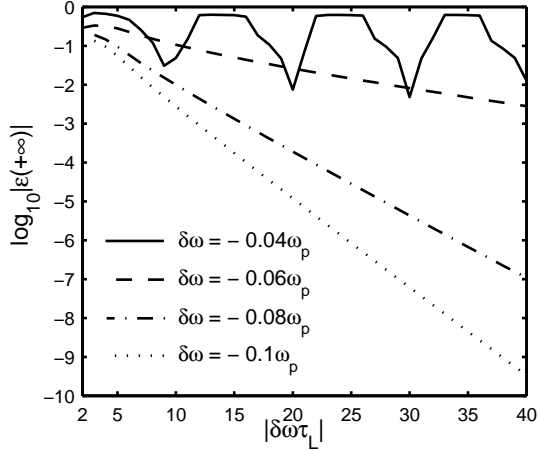}
\caption{\label{AppB1} Amplitude of the plasma wake excited by the
detuned beatwave as a function of laser pulse duration;
$a^2=0.04$.}
\end{figure}

Figure~\ref{AppB} illustrates the evolution of the EPW amplitude
for the resonant ($\Omega=\omega_p$) and near-resonant
($\Omega=0.9\omega_p$) excitation. In the near-resonant case, the
beatwave pulse amplitude varies slowly:
$|\delta\omega\tau_L|\approx30$. Resonant plasma beatwave
[numerical solution of Eq.~(\ref{B1})] exhibits oscillations due
to the periodic dephasing produced by the relativistic frequency
shift~\cite{rosenbluth_liu,Tang}. This effect is eliminated by
introducing the detuning
$\delta\tilde{\omega}\approx-3\Delta=-0.1$ (compare dashed and
solid curves in Fig.~\ref{AppB}). Under parameters of
Fig.~{\ref{AppB}}, $\delta\tilde{\omega}$ dominates the
relativistic frequency shifts [$(3/16)|\varepsilon(t)|^2<0.002$
and $|a_0(t)/2|^2\le0.01$], and the numerical solution of
Eq.~(\ref{B1}) almost coincides with the analytic
approximation~(\ref{B2}).

\begin{figure}[t]
\includegraphics[scale=1]{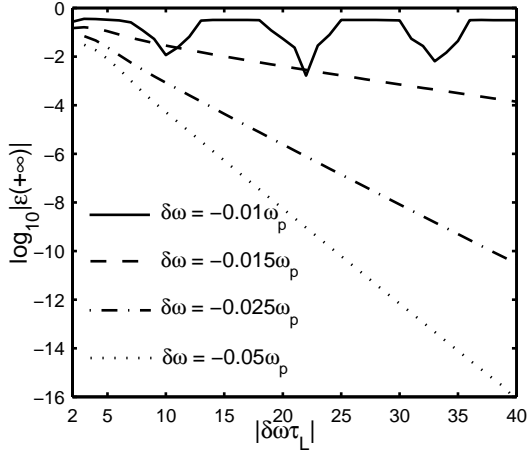}
\caption{\label{AppB2} Amplitude of the plasma wake excited by the
detuned beatwave as a function of laser pulse duration;
$a^2=0.005$.}
\end{figure}

The amplitude of the seed plasma wave from which the SFRS can grow
inside plasma can be estimated using the the amplitude of wake
plasma wave left behind the laser, $|\varepsilon(t\to+\infty)|$.
We evaluate it numerically from Eq.~(\ref{B1}) and plot as a
function of $\tau_L$ for various negative detunings $\delta\omega$
in Figs.~\ref{AppB1} (with the maximal laser intensity as of
Fig.~\ref{Fig2}) and Fig.~\ref{AppB2} (with the maximal laser
intensity as of Fig.~\ref{Fig4}). For the large negative detunings
($\delta\omega<-0.055\omega_p$ for Fig.~\ref{AppB1}, and
$\delta\omega<-0.0125\omega_p$ for Fig.~\ref{AppB2}) the plasma
wake amplitude drops with $\tau_L$ monotonously, while for smaller
magnitudes of $|\delta\omega|$ (numerical examples are given for
$\delta\omega=-0.04\omega_p$ in Fig.~\ref{AppB1}, and for
$\delta\omega=-0.01\omega_p$ in Fig.~\ref{AppB2}) it exhibits an
oscillatory behavior. As the pulse duration $\tau_L$ grows, the
wake amplitude periodically reaches the value
$|\varepsilon(t\to+\infty)|\approx0.7$ (Fig.~\ref{AppB1}) and
$|\varepsilon(t\to+\infty)|\approx0.5$ (Fig.~\ref{AppB2}). This is
the signature of relativistic bi-stability of the
system~\cite{bistability,Royal}.




\end{document}